\documentclass[aps,prb,10pt, twocolumn, superscriptaddress, longbibliography, nobalancelastpage, nofootinbib]{revtex4-2}

\usepackage{graphicx}
\usepackage{graphics}
\usepackage{amsmath}
\usepackage{amssymb}
\usepackage{amsfonts}
\usepackage{dsfont}
\usepackage{braket}
\usepackage{color}
\usepackage{braket,slashed}
\usepackage[mathscr]{euscript}
\definecolor{darkblue}{rgb}{0, 0, 0.8}
\usepackage[colorlinks=true, breaklinks=true, linkcolor=red, citecolor=blue, urlcolor=blue]{hyperref} 
\usepackage{hyperref}
\usepackage{subfigure}
\usepackage{xfrac}
\usepackage{bm}
\usepackage{kantlipsum}
\usepackage{enumitem}
\usepackage{tikz}
\usepackage{framed}
\usepackage{graphicx}
\usepackage{subfigure}
\usepackage{cleveref}
\usepackage{csquotes}

\allowdisplaybreaks[1]

\newcommand{\code}[1]{\texttt{#1}}

\newcommand{\bM}{\ensuremath{{\bar{M}}}}
\newcommand{\mT}{\ensuremath{{\mathcal{T}}}}

\DeclareMathOperator*{\argmin}{arg\ min}
\DeclareMathOperator*{\argmax}{arg\ max}

\renewcommand{\dag}{^\dagger}

\newcommand{\e}{\ensuremath{\mathrm{e}}}

\newcommand{\SU}{\ensuremath{\mathrm{SU}}}

\newcommand{\U}{\ensuremath{\mathrm{U}}}

\newcommand{\diagram}[2]{\;\vcenter{\hbox{\includegraphics[scale=0.32,page=#2]{diagram_#1.pdf}}}\;}

\newcommand{\cP}{\ensuremath{\mathcal{P}}}
\newcommand{\cZ}{\ensuremath{\mathcal{Z}}}
\newcommand{\txe}{\ensuremath{\text{e}}}
\newcommand{\diagramScale}[3]{\;\vcenter{\hbox{\includegraphics[scale=#3,page=#2]{diagram_#1.pdf}}}\;}

\begin{document}

\title{Variational methods for contracting projected entangled-pair states}

\newcommand{\gent}[0]{Department of Physics and Astronomy, University of Ghent, Krijgslaan 281, 9000 Gent, Belgium}
\newcommand{\amsterdam}[0]{Institute for Theoretical Physics and Delta Institute for Theoretical Physics, University of Amsterdam, Science Park 904, 1098 XH Amsterdam, The Netherlands}
\newcommand{\vienna}[0]{Faculty of Physics, University of Vienna, Boltzmanngasse 5, 1090 Wien, Austria}

\author{Laurens Vanderstraeten}
\email{laurens.vanderstraeten@ugent.be}
\affiliation{\gent}
\author{Lander Burgelman}
\affiliation{\gent}
\author{Boris Ponsioen}
\affiliation{\amsterdam}
\author{Maarten Van Damme}
\affiliation{\gent}
\author{Bram Vanhecke}
\affiliation{\gent}
\affiliation{\vienna}
\author{Philippe Corboz}
\affiliation{\amsterdam}
\author{Jutho Haegeman}
\affiliation{\gent}
\author{Frank Verstraete}
\affiliation{\gent}

\begin{abstract}
The norms or expectation values of infinite projected entangled-pair states (PEPS) cannot be computed exactly, and approximation algorithms have to be applied. In the last years, many efficient algorithms have been devised -- the corner transfer matrix renormalization group (CTMRG) and variational uniform matrix product state (VUMPS) algorithm are the most common -- but it remains unclear whether they always lead to the same results. In this paper, we identify a subclass of PEPS for which we can reformulate the contraction as a variational problem that is algorithm independent. We use this variational feature to assess and compare the accuracy of CTMRG and VUMPS contractions. Moreover, we devise a new variational contraction scheme, which we can extend to compute general $N$-point correlation functions.
\end{abstract}

\maketitle

\section{Introduction}

Tensor networks provide excellent variational states for approximating strongly-correlated ground states of generic quantum lattice systems. This can be brought back to their special entanglement properties, which are characteristic for low-energy states of local Hamiltonians. The class of matrix product states (MPS) \cite{Schollwoeck2011} and projected entangled-pair states (PEPS) \cite{Verstraete2004} are the most versatile examples for simulating one-dimensional (1-D) and two-dimensional (2-D) systems, respectively. They have a built-in control parameter, the bond dimension $D$. For gapped systems tensor networks are believed to yield essentially exact ground states as long as the bond dimension is chosen large enough \cite{Verstraete2006}, whereas for gapless systems a finite-entanglement scaling gives access to the critical data \cite{Nishino1996b, Tagliacozzo2008, Pollmann2009, Pirvu2012, Vanhecke2019, Rader2018, Corboz2018, Czarnik2019, Vanhecke2021}.

\par For tensor networks to be useful for simulating realistic models, we need to be able to compute expectation values and perform variational optimizations efficiently. In the case of MPS, the density-matrix renormalization group \cite{White1992} and its extensions \cite{Schollwoeck2011} provide efficient numerical schemes that are used for simulating 1-D quantum systems with high accuracy. For PEPS, the situation is less clear. First of all, the computation of expectation values can only be done approximately \cite{Schuch2007} and often requires large computational resources. Here a second control parameter enters, the environment bond dimension $\chi$. There are different algorithms for performing this approximate evaluation (see below), and it is a priori not clear that they give identical results for the same value of $\chi$. Secondly, the different algorithms for optimizing the PEPS tensors can give rise to different results: The most widely-used PEPS algorithms rely on imaginary-time evolution with local truncations, where either the environment is approximated in a mean-field fashion (simple update) \cite{Jiang2008} or fully taken into account (full update) \cite{Jordan2008}. It is only by the advent of variational optimization methods for PEPS \cite{Corboz2016, Vanderstraeten2016} that it was realized that even the full-update algorithm does not necessarily find the optimal PEPS tensors. As a result, it proved crucial to use PEPS results from variational optimization when extrapolating finite-$D$ results for critical models \cite{Rader2018, Corboz2018, Vanhecke2021}.

\par In this paper, we show that under certain symmetry conditions on the PEPS tensors, the problem of contracting a PEPS can be turned into a variational problem itself. This implies that we can determine the approximate energy expectation value, as well as other observables, for a given value of the environment bond dimension $\chi$, independently of the algorithm that is actually used for computing it. This solves some of the ambiguities in the practice of PEPS optimization, since it implies that we can formulate a variational principle for a PEPS class of states at a certain bond dimension by specifying the environment bond dimension $\chi$. Moreover, this variational characterization allows us to use variational MPS algorithms for PEPS contractions and evaluating correlation functions. Finally, it allows us to compare the accuracy of different, possibly non-variational, contraction schemes.

\par The paper is organized as follows. We first review (Sec.~\ref{sec:review}) the different contraction algorithms that are currently used in PEPS simulations, and motivate the use of an algorithm-independent characterization of the environment. In the next section (Sec.~\ref{sec:symmetries}), we give symmetry constraints on the PEPS such that the transfer matrix becomes Hermitian. Then (Sec.~\ref{sec:contraction}) we show that this allows us to formulate a variational principle for the PEPS contraction, and we describe variational algorithms for finding the environment. We go on (Sec.~\ref{sec:summation}) with a new algorithm for computing two-point functions, using MPS techniques that resemble state-of-the-art methods for evaluating spectral functions in quantum spin chains. In Sec.~\ref{sec:benchmarks} we benchmark these variational contraction and summation algorithms, and compare with other established methods.

\section{Review of contraction algorithms}
\label{sec:review}

Let us first review the PEPS construction, and the most common contraction algorithms. A PEPS can be directly defined on an infinite lattice by repeating a unit cell of tensors. In the simplest case, the unit cell consists of just a single tensor $A$ and the state can be represented diagrammatically as\footnote{Here it is understood that the PEPS tensor network encodes the expansion coefficients of the state in the basis of spin configurations.}
\begin{equation} \label{eq:peps_single}
\ket{\Psi(A)} = \diagram{p2_review}{1}.
\end{equation}
The four virtual indices of the PEPS tensors are contracted, whereas the physical index represents the physical degree of freedom. The state is thus parametrized by a single five-leg tensor $A$, where the bond dimension $D$ of the four virtual indices can be increased to enlarge the variational class of states.
\par The first step in dealing with these infinite PEPSs consists of normalizing the state, i.e. we want to compute the norm
\begin{equation} \label{eq:norm}
n(A,\bar{A}) = \braket{\Psi(\bar{A}) | \Psi(A) }
\end{equation}
directly in the thermodynamic limit. We can represent this norm as an infinite tensor network
\begin{equation}
n(A,\bar{A}) = \diagram{p2_review}{2},
\end{equation}
with the double-layer tensor $O$ obtained by the contraction of the tensor $A$ with its conjugate
\begin{equation}
\diagram{p2_review}{3} = \diagram{p2_review}{4}.
\end{equation}
Similarly, a local expectation value is represented as
\begin{equation}
\diagram{p2_review}{5},
\end{equation}
where the dashed lines indicates the location of a local operator squeezed between the two layers.

\par In practical PEPS algorithms, there are two approaches for computing such infinite tensor networks. The first one is based on the corner-transfer matrix renormalization group (CTMRG) \cite{Baxter1968, Baxter1978, Nishino1996, Nishino1997, Orus2009, Corboz2010, Corboz2014}, which finds a local environment in the form of
\begin{equation} \label{eq:ctmrg1}
\diagram{p2_review}{6},
\end{equation}
where the $C$ and $T$ tensors\footnote{For generic PEPSs, the four $C$ and $T$ tensors are, in fact, different.} represent the infinite network surrounding this site; the dimension of the legs of these tensors is $\chi$. The environment tensors are found numerically by iterating a real-space renormalization step until convergence: in each step ($i$) an $O$ tensor is absorbed into the environment tensors and ($ii$) the bond dimension of the tensors is truncated to $\chi$. There are different prescriptions for this truncation step, and they vary both in stability of the algorithm and accuracy of the results. The expectation value of a larger local operator is computed by enlarging the same environment in a trivial way, e.g.
\begin{equation} \label{eq:ctmrg2}
\diagram{p2_review}{7}.
\end{equation}

\par The second approach relies on finding approximate MPS for representing the fixed point of the one-dimensional transfer matrix,
\begin{equation}
\mT = \diagram{p2_review}{8}.
\end{equation}
This fixed-point equation is written as
\begin{equation}
\diagram{p2_review}{9} \approx \diagram{p2_review}{10}
\end{equation}
where $\chi$ is now the bond dimension of the boundary MPS. Different algorithms can be used to find the approximate fixed point such as the density-matrix renormalization group \cite{Nishino1995}, a variant of the time-evolving block decimation \cite{Orus2008} or the variational uniform MPS (VUMPS) algorithm \cite{Vanderstraeten2015, Haegeman2017, Fishman2018, Vanderstraeten2019a}. Once such a boundary MPS is found from both directions parametrized by $M$ and $\tilde{M}$ tensors\footnote{In general there is no simple relation between $M$ and $\tilde{M}$, but in many cases the spatial symmetries of the PEPS allow to relate them in a simple way (see below).}, the local environment is found as
\begin{equation}
\diagram{p2_review}{11}.
\end{equation}
Again, larger environments can be found straightforwardly.
\par Evaluating non-local expectation values such as correlation functions can be done as well. For a two-point correlation function on the same row or column in the lattice, this is easily done by extending CTMRG environment,
\begin{equation}
\diagram{p2_review}{12} \cdots \diagram{p2_review}{13}
\end{equation}
or taking the boundary MPS
\begin{equation}
\diagram{p2_review}{14} \cdots \diagram{p2_review}{15}.
\end{equation}
More general two-point correlation functions can be computed by extending boundary MPS to include corners, giving rise to corner-shaped environments \cite{Vanderstraeten2015} or by a summation routine based on the CTMRG scheme \cite{Corboz2016}.
\par This review makes clear that there is a plethora of different types of environments and algorithms for contracting a given PEPS with a certain environment bond dimension $\chi$. It is expected that they give different results for computing local expectation values, and it is not even clear that they all converge to the same result in the $\chi\to\infty$ limit. In the following sections, we aim at partially solving these ambiguities.

\section{Symmetries and the transfer matrix}
\label{sec:symmetries}

Imposing symmetry constraints on the PEPS tensors is a powerful tool for restricting the variational space and making algorithms more efficient. A first class of constraints involves on-site symmetries of the state: Imposing that a PEPS wavefunction is invariant under a global on-site symmetry action can be imposed on the level of the local PEPS tensor. This way of constraining the PEPS tensor has been used successfully to simulate ground states with (abelian) $\U(1)$ symmetries \cite{Bauer2011, Hasik2020} and (non-abelian) $\SU(2)$ \cite{Mambrini2016, Poilblanc2017, Chen2018} or $\SU(N)$ \cite{Kurecic2019, Gauthe2019, Chen2020} symmetries.
\par Here we are interested in a second class of constraints, involving spatial symmetries of the PEPS wavefunction. Again, we can impose symmetry constraints on the level of the local PEPS tensor. One type of spatial symmetry that is often useful is rotation invariance of the PEPS tensor, 
\begin{equation} \label{eq:rotation}
\diagram{p3_symm}{2}=\diagram{p3_symm}{3},
\end{equation}
leading to a global state that is rotation invariant. A second spatial symmetry is reflection, which is typically combined with time-reversal symmetry,
\begin{equation} \label{eq:PT}
\diagram{p3_symm}{4} = \diagram{p3_symm}{5} = \diagram{p3_symm}{6},
\end{equation}
where the bar denotes complex conjugation of the tensor elements. Note that an explicit breaking of reflection and time-reversal symmetry separately often leads to chiral wavefunctions \cite{Poilblanc2016, Hackenbroich2018}. The interplay between internal symmetries and spatial symmetries leads to non-trivial classes of PEPS wavefunctions \cite{Jiang2015}.
\par Here we are interested in the latter symmetry, and in the relation with the transfer matrix. Indeed, if we take the Hermitian conjugate of the transfer matrix
\begin{equation}
\mT\dag = \diagram{p3_symm}{7},
\end{equation}
we see that this is invariant when the symmetry condition in Eq.~\ref{eq:PT} is imposed on the $A$ tensor. So this symmetry implies that the transfer matrix is Hermitian, but the Hermiticity requirement can be met by a more general condition. Indeed, under the condition
\begin{equation}
\diagram{p3_symm}{8} = \diagram{p3_symm}{9}
\end{equation}
with $U$ a unitary matrix and $X$ an invertible matrix, the transfer matrix is hermitian by construction.
\par This hermiticity condition can be extended to larger unit cells, where we require that the transfer matrix of a full unit cell is hermitian. This requirement can be illustrated in the case of a PEPS with a two-by-two unit cell, 
\begin{equation}
\ket{\Psi(A_i)} = \diagram{p3_symm}{10}.
\end{equation}
Indeed, if we require
\begin{align}
\diagram{p3_symm}{11} &= \diagram{p3_symm}{12} \\ 
\diagram{p3_symm}{13} &= \diagram{p3_symm}{14}
\end{align}
and similar conditions on $A_3$ and $A_4$, then the two-row transfer matrix
\begin{equation}
\mT = \diagram{p3_symm}{15}
\end{equation}
is Hermitian.

\section{Variational contraction}
\label{sec:contraction}

In the previous section we have explained under what conditions the PEPS transfer matrix is Hermitian. In this section, we show why this feature is extremely useful for characterizing and finding a boundary MPS.

\subsection{Single-row transfer matrix}

The norm of a PEPS $n(A,\bar{A})$ can be interpreted as an infinite product of transfer matrices and, therefore, reduces to an infinite power of the leading eigenvalue. The corresponding eigenvalue equation is given by
\begin{equation} \label{eq:eigEq}
\mT \ket{\Psi} = \Lambda \ket{\Psi},
\end{equation}
where $\Lambda$ scales exponentially with the number of sites in the $x$-direction, i.e. $\Lambda \propto \lambda^{N_x}$. In analogy with transfer-matrix approaches in statistical mechanics \cite{Haegeman2017}, we can associate a free-energy density to the norm of the infinite PEPS as
\begin{align}
f(A,\bar{A}) &= - \lim_{N_x,N_y\to\infty} \frac{1}{N_xN_y} \log n(A,\bar{A}) \nonumber \\
&= - \log \lambda .
\end{align}
Since the transfer matrix is a Hermitian operator, the above eigenvalue equation can be reformulated as a variational problem
\begin{equation} \label{eq:costFun}
\ket{\Psi} =  \argmin_{\ket{\Psi}} \left( - \frac{1}{N_x} \log \left( \frac{\bra{\Psi} \mT \ket{\Psi}} {\braket{\Psi|\Psi}} \right) \right).
\end{equation}
We can now solve the eigenvalue equation variationally by approximating the fixed point as an infinite MPS, parametrized by a single tensor $M$ 
\begin{equation}
\ket{\Psi_M} = \cdots \diagram{p4_contr1}{2} \cdots.
\end{equation}
The variational characterization of the MPS tensor is given by the following variational optimization problem
\begin{equation} \label{eq:var_mps1}
M = \argmin_{M} \left( - \frac{1}{N_x} \log \Lambda(M,\bar{M}) \right)
\end{equation}
with
\begin{equation}
\Lambda(M,\bar{M}) =  \left( \frac{\bra{\Psi_\bM} \mT \ket{\Psi_M}} {\braket{\Psi_\bM|\Psi_M}} \right) .
\end{equation}
Diagrammatically, the numerator is represented as an infinite channel
\begin{equation}
\bra{\Psi_\bM} T \ket{\Psi_M} = \cdots \diagram{p4_contr1}{3} \cdots,
\end{equation}
which scales as $\Lambda\propto\lambda^{N_x}$. The value $\lambda$ is determined as the leading eigenvalue of the channel operator
\begin{equation}
\lambda = \rho_\mathrm{max} \left( \diagram{p4_contr1}{4} \right).
\end{equation}
\par Given this variational characterization of the boundary MPS, we can use gradient-based optimization techniques to find it. The gradient of the objective function in Eq.~\eqref{eq:var_mps1} is given by
\begin{align}
g &= 2\partial_{\bM} \left( - \frac{1}{N_x} \log \Lambda(M,\bM) \right) \nonumber \\
&= - \frac{2}{N_x} \frac{1}{\Lambda} \frac{ \bra{\partial_\bM\Psi_\bM} \left( \mT - \Lambda \right) \ket{\Psi_M}} {\braket{\Psi_\bM|\Psi_M}} .
\end{align}
Diagrammatically, the gradient is given by
\begin{multline} \label{eq:gradient_mps1}
g = -2 \left( \frac{1}{\lambda} \cdots \diagram{p4_contr1}{5} \cdots \right. \\ \left.  \phantom{\diagram{p4_contr1}{5}} - \cdots \diagram{p4_contr1}{6} \cdots \right),
\end{multline}
assuming the MPS is normalized. We can now use efficient optimization techniques over MPS manifolds \cite{Hauru2020} to optimize the boundary MPS, exploiting the gauge degrees of freedom in the MPS ansatz.
\par Instead of using gradient-based techniques, we can reformulate the variational characterization in Eq.~\ref{eq:var_mps1} as a fixed-point equation, and, starting from a random MPS, iterate this equation until it converges to the variational optimum. This procedure is known as the VUMPS algorithm \cite{ZaunerStauber2018, Fishman2018}. First, we bring the MPS into center-gauged form as
\begin{equation}
\ket{\Psi_M} = \cdots \diagram{p4_contr1}{7} \cdots,
\end{equation}
where $M^l$ and $M^r$ are related to the original MPS tensor via a gauge transform and satisfy the left- and right-canonical gauge condition
\begin{equation} \label{eq:vumps1}
\diagram{p4_contr1}{8} = \diagram{p4_contr1}{9}, \qquad \diagram{p4_contr1}{10} =  \diagram{p4_contr1}{11},
\end{equation}
and are related via the matrix $C$
\begin{equation} \label{eq:vumps2}
\diagram{p4_contr1}{12} = \diagram{p4_contr1}{13} = \diagram{p4_contr1}{14}.
\end{equation}
In this MPS representation, the condition of a vanishing gradient is equivalent to the fixed-point equation for $M^c$
\begin{equation} \label{eq:vumps3}
\cdots \diagram{p4_contr1}{15} \cdots = \lambda \diagram{p4_contr1}{16},
\end{equation}
and a similar equation for the matrix $C$ is easily derived:
\begin{equation} \label{eq:vumps4}
\cdots \diagram{p4_contr1}{17} \cdots = \lambda \diagram{p4_contr1}{18}.
\end{equation}
These equations [Eqs.~\eqref{eq:vumps1}-\eqref{eq:vumps4}] fully characterize the variationally optimal boundary MPS; in Refs.~\onlinecite{Fishman2018} and \onlinecite{ Vanderstraeten2019a} the iterative scheme is explained in detail.

\subsection{Two-row transfer matrix}

Suppose we have a PEPS with a two-by-two unit cell (the more general case of a two-by-$N$ unit cell is treated similarly), giving rise to a two-row transfer matrix $\mT=\mT_2 \mT_1$. As explained above, we parametrize the PEPS such that the two-row transfer matrix is Hermitian, i.e. we have $\mT_2\dag =\mT_1$. We characterize the fixed point of this two-row transfer matrix as
\begin{align}
& \mT_1 \ket{\Psi_1} = \Lambda_1 \ket{\Psi_2} \nonumber \\
& \mT_2 \ket{\Psi_2} = \Lambda_2 \ket{\Psi_1}.
\end{align}
in terms of two normalized states $\ket{\Psi_1}$ and $\ket{\Psi_2}$. Under the Hermiticity condition, $\ket{\Psi_1}$ can be characterized variationally as the state which optimizes the eigenvalue of the two-row transfer matrix
\begin{equation} 
\ket{\Psi_1} = \argmin_{\ket{\Psi_1}} \left( - \frac{1}{N_x} \log \frac{\bra{\Psi_1} \mT_2\mT_1 \ket{\Psi_1}  } {\braket{\Psi_1|\Psi_1} }\right).
\end{equation}
Here, however, we can go one step further. The eigenvectors $\ket{\Psi_1}$ and $\ket{\Psi_2}$ are, in fact, the left and right leading singular vectors of the matrix $\mT_1$, and we can use the variational characterization for singular vectors to reformulate the optimality condition as
\begin{equation} \label{eq:var_mps2}
\ket{\Psi_1},\ket{\Psi_2} = \argmin_{\ket{\Psi_1},\ket{\Psi_2}} \left( - \frac{1}{N_x} \log\Lambda \right),
\end{equation}
with
\begin{equation}
\Lambda = \frac{\bra{\Psi_2} \mT_1 \ket{\Psi_1} \bra{\Psi_1} \mT_2 \ket{\Psi_2} } {\braket{\Psi_1|\Psi_1} \braket{\Psi_2|\Psi_2}}.
\end{equation}
The Hermiticity condition $\mT_2\dag=\mT_1$ ensures that this cost function is real.
\par We can now approximate the fixed point as a set of two MPSs, which we also take to have a two-site unit cell and are, therefore, each described by two MPS tensors
\begin{equation}
\ket{\Psi_1} \approx \ket{\Psi_{M_{11},M_{12}}}, \qquad
\ket{\Psi_2} \approx \ket{\Psi_{M_{21},M_{22}}}.
\end{equation}
The above cost function can be expressed as (with $\Lambda_1=\bar{\Lambda}_2$)
\begin{equation}
\Lambda = \Lambda_1 \Lambda_2, 
\end{equation}
with
\begin{align}
& \Lambda_1 = \cdots \diagram{p4_contr2}{1} \cdots \nonumber \\
&\Lambda_2 = \cdots \diagram{p4_contr2}{2} \cdots.
\end{align}
Again, we can use optimization techniques over MPS manifolds to optimize these two MPSs efficiently.
\par Analoguously to the above case of a one-row transfer matrix, we can switch to a center-gauge representation for the MPSs and derive fixed-point equations for the MPS tensors. Here, we find 
\begin{equation}
\cdots \diagram{p4_contr2}{3} \cdots = \lambda_{11} \diagram{p4_contr2}{4}
\end{equation}
for the center-site MPS tensors (the other three equations follow similarly). These equations constitute the multi-site version of the VUMPS algorithm, which was explained in detail in Ref.~\onlinecite{Nietner2020}. Starting from a set of random MPSs, the VUMPS fixed-point equations can be iterated in order to find the optimal MPSs. 

\subsection{More than two rows}
\label{sec:vumps_mps3}

Let us now consider the case of a three-row transfer matrix, with the Hermiticity conditions $\mT=\mT_1\dag \mT_2 \mT_1$ and $\mT_2\dag = \mT_2$. The fixed point is given by
\begin{align} \label{eq:3eq}
\mT_1 \ket{\Psi_1} &= \Lambda_1 \ket{\Psi_2} \nonumber \\
\mT_2 \ket{\Psi_2} &= \Lambda_2 \ket{\Psi_3} \nonumber \\
\mT_1\dag \ket{\Psi_3} &= \Lambda_3 \ket{\Psi_1},
\end{align}
and the state $\ket{\Psi_1}$ can be characterized variationally as
\begin{equation} 
\ket{\Psi_1} = \argmin_{\ket{\Psi_1}} \left( - \frac{1}{N_x} \log \frac{\bra{\Psi_1} T_1\dag T_2 T_1 \ket{\Psi_1}  } {\braket{\Psi_1|\Psi_1} }\right).
\end{equation}
Here, however, we \emph{cannot} take the step of reformulating this as a variational principle where we optimize all layers simultaneously,
\begin{equation} \label{eq:var_mps3}
\ket{\Psi_1},\ket{\Psi_2},\ket{\Psi_3} \neq \argmin_{\ket{\Psi_1},\ket{\Psi_2},\ket{\Psi_3}} \left( - \frac{1}{N_x} \log\Lambda \right)
\end{equation}
with
\begin{equation} \label{eq:costFun_mps3}
\Lambda = \frac{\bra{\Psi_2} \mT_1 \ket{\Psi_1} \bra{\Psi_3} \mT_2 \ket{\Psi_2} \bra{\Psi_1} \mT_1\dag \ket{\Psi_3}} {\braket{\Psi_1|\Psi_1} \braket{\Psi_2|\Psi_2} \braket{\Psi_3|\Psi_3}},
\end{equation}
as this cost function is not real, since it is not guaranteed that $ \ket{\Psi_3}$ and $ \ket{\Psi_2} $ are proportional.
\par In order to still obtain a boundary MPS algorithm, we instead start from the set of equations in Eq.~\eqref{eq:3eq}. Assuming the first state can be approximated by an MPS, $\ket{\Psi_1} \approx \ket{\Psi_{\{M_1\}}}$, we can find an MPS for the second state by applying the operator $\mT_1$ to $\ket{\Psi_{\{M_1\}}}$ and approximating the result by an MPS with smaller bond dimension. This approximation $\ket{\Psi_{\{M_2\}}}$ can, again, be characterized by a variational principle, where we now optimize the (normalized) fidelity
\begin{equation}
\{M_2\} = \argmax_{\{M_2\}}  \frac{ \left|\bra{\Psi_{\{M_2\}}} \mT_1 \ket{ \Psi_{\{M_1\}}} \right|^2} {\braket{\Psi_{\{M_2\}}|\Psi_{\{M_2\}}}},
\end{equation}
and we have similar equations for the $\{M_1\}$ and $\{M_3\}$ tensors. In Ref.~\onlinecite{Vanhecke2021b} it was shown that these optimization problems can be reformulated in center-gauge representation to obtain fixed-point equations characterizing each set of MPS tensors $\{M_i\}$. This naturally leads to a variational boundary MPS algorithm consisting of a power method with a variational optimization of the individual fidelities in each step.
\par If one instead solves all these fixed-point equations simultaneously by optimizing over all MPS at once, this gives rise to a single set of equations which is similar to the ones we have given for one- and two-row transfer matrices \cite{Nietner2020}. Thus, by resorting to a method which optimizes all eigenvalue equations [Eq.~\ref{eq:3eq}] simultaneously, we essentially find the VUMPS fixed-point equations that would follow had Eq.~\ref{eq:var_mps3} been a valid variational principle. Given that this is not the case, and there is no variational principle underlying those fixed point equations, we expect that iterating the multi-site VUMPS fixed-point equations will, in general, not always work for transfer matrices with three or more rows, even if the total transfer operator is Hermitian. In Appendix \ref{sec:multivumps}, we provide a detailed diagnostic of this issue. 
\par We conclude by noting that the computational complexity of this multi-site VUMPS algorithm scales linearly in $L_x$ and $L_y$, where $L_x$ is the MPS unit-cell size and $L_y$ is the number of rows in the transfer matrix.

\section{Summation of two-point functions}
\label{sec:summation}

Besides the norm and local expectation values, non-local two-point functions -- or $N$-point functions, in general -- are often relevant observables of a given PEPS wavefunction. In Ref.~\onlinecite{Vanderstraeten2015} an extension of the boundary-MPS approach was introduced for computing the most general two-point and three-point functions, and in Ref.~\onlinecite{Corboz2016} the CTMRG approach was extended to perform the same task; finally, in Ref.~\onlinecite{WeiLin2021} a method was introduced for summing two-point functions using generating functions. In this section, we propose a different scheme that works for generic $N$-point functions by translating standard MPS algorithms to this context.

\subsection{Static structure factor}

Let us focus on a generic momentum-resolved two-point function, or static structure factor, of the form
\begin{equation}
s(\vec{k}) = \sum_{mn} \e^{i(k_xm + k_yn)}  \bra{\Psi(A)} S_{m,n} S_{0,0} \ket{\Psi(A)},
\end{equation}
with $S_{m,n}$ a local operator acting on site $(m,n)$, which requires resumming an infinite number of two-point correlators. We assume that we have found a normalized boundary MPS, which we can bring into center-gauged form
\begin{equation}
\ket{\Psi_M} = \cdots \diagram{p5_structure}{1},
\end{equation}
and that we have normalized the PEPS.
\par The first contribution to the structure factor is the one where both operators act on the same site, which can be easily computed as a local expectation value
\begin{equation}
s_0 = \cdots \diagram{p5_structure}{2} \cdots,
\end{equation}
where, as before, the dashed lines indicate the presence of the local operators.
We can introduce the left- and right fixed points of the channel operator,
\begin{equation} \label{eq:GL}
\diagram{p5_structure}{3}=\diagram{p5_structure}{4},
\end{equation}
and
\begin{equation} \label{eq:GR}
\diagram{p5_structure}{5}=\diagram{p5_structure}{6},
\end{equation}
such that this expectation value is given by the network contraction
\begin{equation}
s_0 = \diagram{p5_structure}{7}.
\end{equation}
Summing contributions where the two operators live on the same row of the lattice is also straightforward. Indeed, all these contributions can be grouped under an expression of the form
\begin{align}
\e^{ik_x} \sum_n  \diagram{p5_structure}{8} \left( \e^{ik_x} \diagram{p5_structure}{9} \right)^n \diagram{p5_structure}{10}.
\end{align}
The infinite series can be taken explicitly, and we find 
\begin{multline}
s_x(k_x) = \e^{ik_x}\diagram{p5_structure}{8} \left( 1 - \e^{ik_x} \diagram{p5_structure}{9} \right)^{-1} \\ \diagram{p5_structure}{10} + \mathrm{h.c.}
\end{multline}
Summing up contributions where the two operators live on different lines is less straightforward. Indeed, these all have the form
\begin{equation}
\diagram{p5_structure}{11}
\end{equation}
with a power of the transfer matrix in between the two operators. If we call the state
\begin{equation}
\ket{\Psi_M(S_i)} = \diagram{p5_structure}{12},
\end{equation}
where the operator acts on site $i$, the above diagram can be denoted as
\begin{equation}
\bra{\Psi_M(S_j)} \mT^{n-1} \ket{\Psi_M(S_i)},
\end{equation}
with $n$ the vertical distance between the two operators. Summing up all these different contributions amounts to
\begin{align}
s_{xy}(\vec{k}) &= \sum_{mn} \e^{i(k_xm+k_yn)} \bra{\Psi_M(S_m)} \mT^{n-1} \ket{\Psi_M(S_{0})} \nonumber \\
&= \sum_m \e^{ik_xm} \bra{\Psi_M(S_m)} \nonumber \\
& \qquad\qquad\qquad  \e^{ik_y}\left(1 - \e^{ik_y} \mT\right)^{-1} \ket{\Psi_M(S_{0})}. \label{eq:sxy}
\end{align}
This expression is very similar to a frequency-resolved dynamical correlation function for a 1-D spin chain, typically of the form
\begin{align}
S(q,\omega) &= \int dt\,  \e^{i\omega t} \sum_m \e^{iqm} \bra{\Psi_0} S_m(t) S_0(0) \ket{\Psi_0} \nonumber \\
&= \sum_m \e^{iqm} \bra{\Psi_0} S_m \left( \omega - (H-E_0) \right)^{-1} S_0 \ket{\Psi_0},
\end{align}
where $\ket{\Psi_0}$ would be the ground state of the spin-chain Hamiltonian $H$ with ground-state energy $E_0$. Many different approaches exist in the MPS literature to compute the latter overlap explicitly, including Lanczos methods \cite{Hallberg1995, Dargel2012} the correction vector approach \cite{Ramasesha1996, Kuhner1999, Jeckelmann2002, Weichselbaum2009a}, Chebyshev expansion \cite{Holzner2011}, etc. A more straightforward approach consists of time evolving the state according to the spin-chain Hamiltonian, and transforming the time-dependent correlator back to frequency space \cite{White2008a, Pereira2008a, Barthel2009, Kjall2011, Seabra2013}. Here, in order to evaluate Eq.~\eqref{eq:sxy}, we take the discretized version of the latter approach, i.e. we take the state $\ket{\Psi_M(O_{0})}$ and sequentially apply the transfer matrix. In order to keep the problem tractable, we represent the state as a window MPS. 

\par A window MPS \cite{Milsted2013, Phien2012, Zauner2015a} is an MPS with a finite window of non-translation-invariant tensors, embedded in a translation-invariant MPS to the left and the right of this window. Such an MPS is given by
\begin{equation}
\ket{\Psi_M(N_i)} = \cdots \diagram{p5_window}{1} \cdots,
\end{equation}
and represents a local perturbation of the infinite MPS.
\par Approximating the initial state as a finite-window MPS can be done in a simple variational step. Indeed, we can approximate it as a window MPS with a single site as
\begin{equation}
\ket{\Psi_M(S_0)} \approx \cdots \diagram{p5_window}{2} \cdots
\end{equation}
where the tensor $N_1$ is found by solving the variational problem
\begin{equation}
N_1^{\mathrm{opt}} = \argmax_{N_1} \frac{ \left| \braket{\Psi_M(N_1) | \Psi_M(S_0)} \right|^2 } { \braket{\Psi_M(N_1) | \Psi_M(N_1)} }.
\end{equation}
Working in the center gauge reduces the norm to the identity
\begin{equation}
\braket{\Psi_M(N_1) | \Psi_M(N_1)} = \diagram{p5_window}{3},
\end{equation}
so that differentiating the above variational objective function gives the solution for $N_1$,
\begin{equation}
N_1^{\mathrm{opt}} = \diagram{p5_window}{4} = \diagram{p5_window}{5}.
\end{equation}
\par We can continue to represent the state
\begin{equation} \label{eq:Tnpsi}
\mT^n \ket{\Psi_M(S_0)}
\end{equation}
as a finite-window MPS by iterating the same procedure. In every step, we start from a window MPS with tensors $\tilde{N}_i$, apply the transfer matrix and approximate the state again as a window MPS with new tensors $N_i$. Every approximation step is done variationally, i.e. we optimize the objective function
\begin{equation}
N_i^{\mathrm{opt}} = \argmax_{N_i} \frac{ | \bra{\Psi_M(N_i) } \mT \ket{\Psi_M(\tilde{N}_i)} |^2 } { \braket{\Psi_M(N_i) | \Psi_M(N_i)} }.
\end{equation}
We can efficiently solve this optimization problem by a sweeping algorithm, where we sequentially optimize the window tensors separately. Optimizing a single tensor $N_i$ is done by bringing the window MPS in center gauge around site $i$, and optimizing as we did above. After we have found a new window MPS representing the state \eqref{eq:Tnpsi}, we can evaluate all contributions to $s_{xy}(\vec{k})$ from that row, and proceed to the next row.
\par In the case of unitary time evolution according to a spin-chain Hamiltonian, one is severely limited due to the growth of entanglement in the time-evolved state. Here this is not the case, since the transfer matrix is an operator with a spectral gap\footnote{The gap in the transfer matrix is directly related to the correlation length of the PEPS. The latter is finite for generic variational PEPS, but there are fine-tuned PEPS wavefunctions that exhibit critical correlations. For the latter PEPSs, a summation scheme is expected to converge more slowly.}; as a result, the evolved state will converge again to the fixed-point MPS after a number of iterations. At that point, the contributions to the structure factor have become negligible, and we can terminate the algorithm. 
\par The size of the window-MPS is a tunable parameter: We expect that the effect of applying the operator will grow in size as layers of the transfer matrix are applied, very similar to the spreading of correlations in the unitary real-time evolution of a quantum spin chain. In Sec.~\ref{sec:benchmarks} we will investigate the effect of the window size on the accuracy.

\subsection{Energy gradient}

The motivation behind the first PEPS summation algorithms \cite{Vanderstraeten2016, Corboz2016} was finding an efficient method for evaluating the gradient of the PEPS energy objective function, which can be used in gradient-based PEPS optimization routines. In the Appendix (App.~\ref{sec:gradient}) we explain in detail how our new summation scheme can be extended for evaluating the gradient of a plaquette interaction term.
\par We should note that, besides summation schemes, the idea of backwards differentiation can be used to evaluate the energy gradient \cite{Liao2019, Hasik2020} efficiently. One important advantage of the latter approach is that it computes the gradient of the energy function at a given value of environment $\chi$, and can therefore be used to perform a gradient optimization of a PEPS with a fixed value of $\chi$. Such a finite-$\chi$ optimization might be interesting to study finite-$\chi$ effects; these can be used for simulating critical systems, where a finite-$\chi$ scaling behaviour can be expected \cite{Vanhecke2021}.
\par In contrast, the approximate summation schemes do not exactly provide the gradient for a given value of $\chi$: only in the infinite-$\chi$ limit can one expect that the energy and gradient evaluation become fully compatible. Therefore, the summation approach can only be used in a PEPS optimization in the infinite-$\chi$ regime -- i.e., the regime where $\chi$ is chosen large enough such that finite-$\chi$ effects are negligible. 

\section{Benchmarks}
\label{sec:benchmarks}

In order to benchmark and compare different contraction schemes, we take optimized PEPS tensors for the $J_1$-$J_2$ model on the square lattice, after a sublattice rotation with $J_2=0$ and $J_2=1/2$, each time with bond dimension $D=5$. After a sublattice rotation, we can represent the ground state by a PEPS with a single-site unit cell. We have imposed both rotation invariance and reflection/time-reversal symmetry [Eqs.~\eqref{eq:rotation}-\eqref{eq:PT}] on the PEPS tensors. The latter symmetry implies that we can use our variational contraction schemes.
\par We have optimized these PEPS tensors by a gradient-based quasi-Newton optimization, where we have applied the gradient evaluation from Sec.~\ref{sec:gradient}. We find energies $e=-0.669376$ and $e=-0.495912$ respectively, which are comparable to the results in Ref.~\onlinecite{Hasik2020}.

\subsection{Variational contraction}

\begin{figure}
\includegraphics[width=\columnwidth]{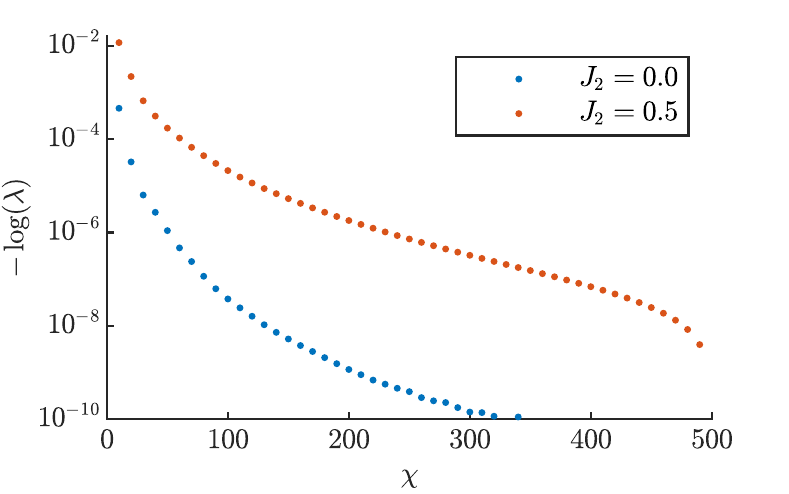}
\caption{The log-norm of the PEPS wavefunction at $J_2=0$ (blue) and $J_2=0.5$ (red) as a function of boundary MPS bond dimension $\chi$. We have rescaled the PEPS tensor such that $\lambda_{\chi=500}=1$ exactly, and do not show results below a cutoff of $10^{-10}$ because they contain numerical noise.}
\label{fig:lambda}
\end{figure}

\begin{figure}
\includegraphics[width=\columnwidth]{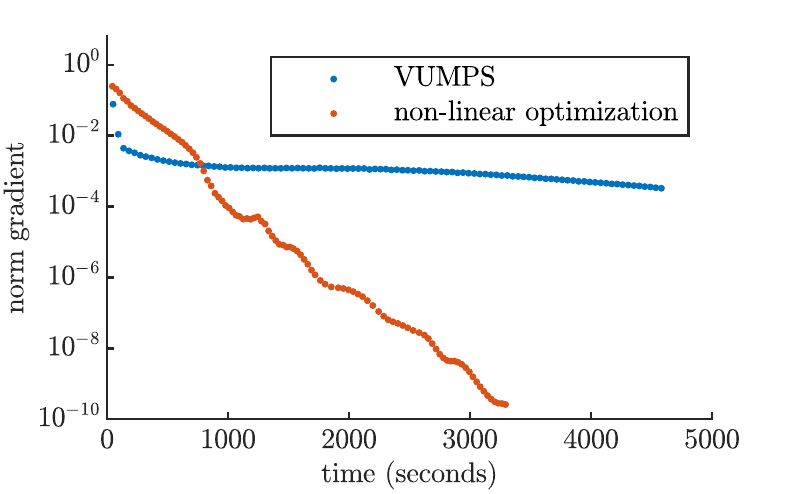}
\caption{The convergence of the boundary MPS with the VUMPS algorithm and by direct variational optimization. We plot the norm of the gradient in Eq.~\eqref{eq:gradient_mps1} for each iteration, as a function of the wall time.}
\label{fig:conv}
\end{figure}

\begin{figure}
\includegraphics[width=\columnwidth]{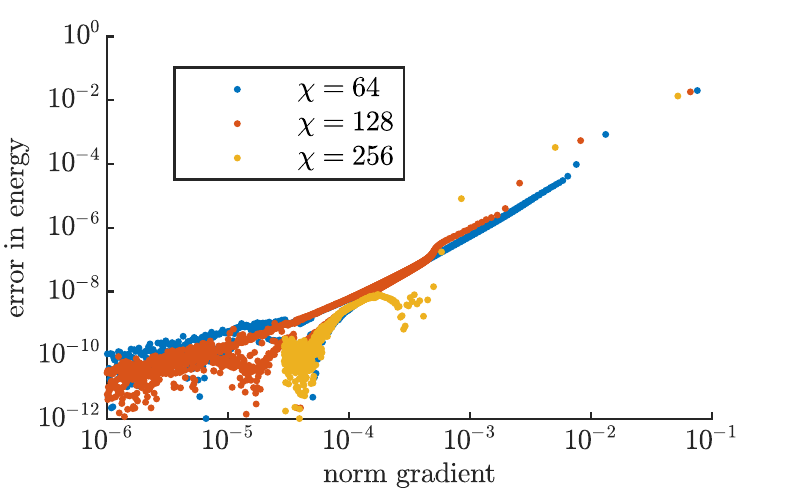}
\caption{The error in evaluating the local energy of the PEPS as a function of the norm of the gradient during the process of optimizing the boundary MPS; we plot this for three different values of $\chi$ for the $J_2=1/2$ PEPS. The error is defined as the absolute difference with the fully optimized boundary MPS.}
\label{fig:energy_vs_error}
\end{figure}

First of all, we explicitly demonstrate the variational nature of the contraction by computing the leading eigenvalue of the PEPS transfer matrix, obtained using an optimized boundary MPS with increasing bond dimension $\chi$. In Fig.~\ref{fig:lambda} one immediately observes that $f=-\log(\lambda)$ monotonically decreases as the bond dimension is increased. The value for the unfrustrated PEPS ($J_2=0$) decreases significantly faster than for the frustrated case ($J_2=1/2$), which points to the fact that the frustrated ground state is more correlated than the unfrustrated one and therefore requires a larger value of $\chi$ for the same accuracy.
\par Next, we assess the performance of a direct optimization of the variational cost function in Eq.~\eqref{eq:var_mps1} by comparing with the VUMPS algorithm. In Fig.~\ref{fig:conv} we plot the norm of the gradient [Eq.~\eqref{eq:gradient_mps1}] after each iteration as a function of the wall time. For the VUMPS algorithm, each iteration consists of (\textit{i}) computing left- and right environments [Eqs.~\eqref{eq:GL}-\eqref{eq:GR}], (\textit{ii}) solving the eigenvalue equations for $M^c$ and $C$ [Eqs.~\eqref{eq:vumps3}-\eqref{eq:vumps4}] and (\textit{iii}) finding new $M^l$ and $M^r$; the two first steps are the most expensive ones. For the direct optimization, we use a quasi-Newton scheme for optimizing over isometric tensor networks \cite{Hauru2020}; each iteration consists of an evaluation of the cost function and the gradient, which again requires computing the left- and right environments [Eqs.~\eqref{eq:GL}-\eqref{eq:GR}]. As compared to VUMPS, the direct optimization strategy does not require solving the eigenvalue equations for $M^c$ and $C$, so each iteration is faster by roughly a factor of 1.5 (the eigenvalue equation for $C$ is not that costly, typically). As VUMPS is a fixed-point method, it can take larger steps in each iteration, allowing to converge much faster initially than a direct-optimization strategy. In the tail of the optimization, however, the manifold optimization clearly outperforms VUMPS fixed-point iterations, and converges more quickly. A hybrid strategy that switches between the two algorithms would be optimal.
\par One might wonder how well the boundary MPS should be converged in order to obtain accurate local PEPS expectation values. In Fig.~\ref{fig:energy_vs_error} we have plotted the error in the energy expectation value for the $J_1-J_2$ Hamiltonian during the process of optimizing the boundary MPS. From this plot, it is clear that pushing the norm of the gradient to very small values is often not necessary in practice.

\subsection{Comparing boundary MPS and CTMRG}

\begin{figure}
\subfigure{\includegraphics[width=\columnwidth]{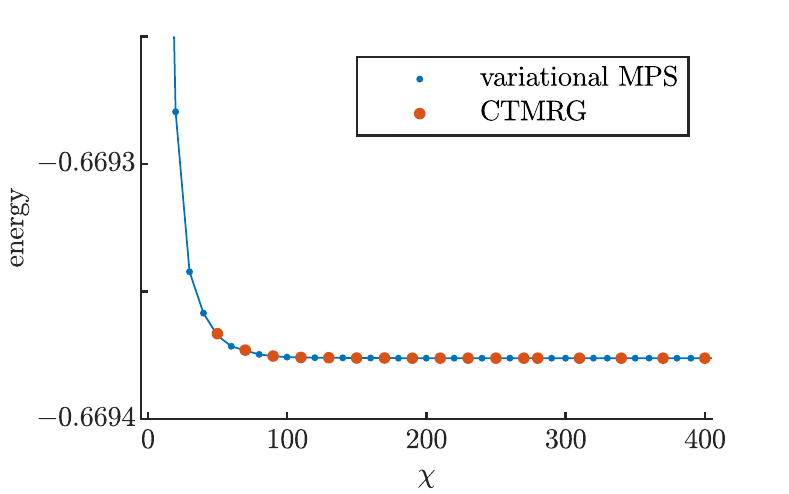}}
\subfigure{\includegraphics[width=\columnwidth]{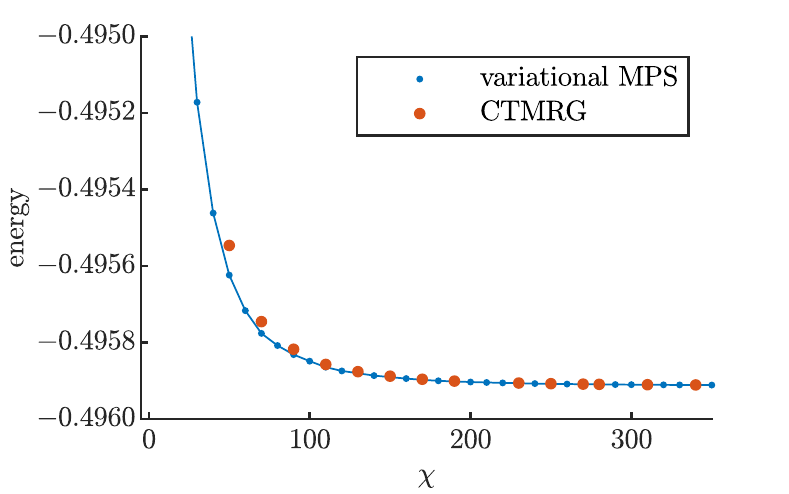}}
\subfigure{\includegraphics[width=\columnwidth]{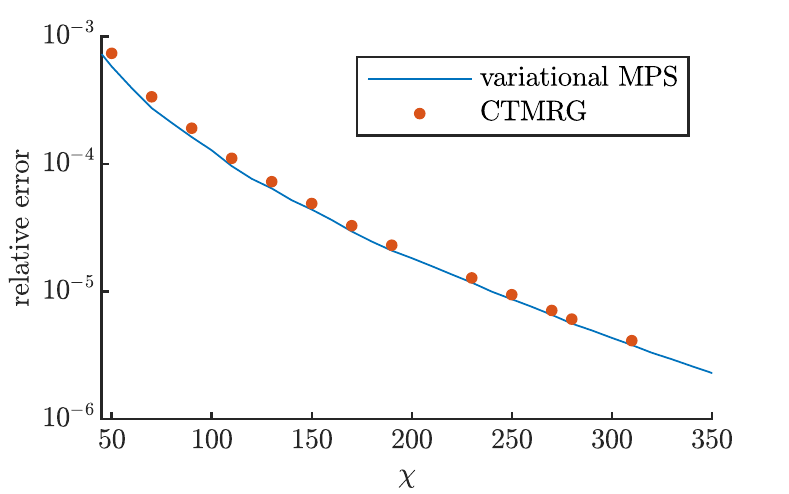}}
\caption{The value for the energy expectation value as a function of $\chi$, computed with the variational MPS and CTMRG environments for $J_2=0$ (top) and $J_2=1/2$ (middle). We also plot the relative error of the energy expectation value at $J_2=1/2$ with respect to the converged value at $\chi=500$ (bottom).}
\label{fig:comp_energy}
\end{figure}

We can now compare the accuracy of boundary-MPS and CTMRG contractions for evaluating local expectation values of PEPS wavefunctions. For the boundary MPS, we use a direct variational optimization as outlined above. For CTMRG, due to the symmetries of the PEPS tensor we can choose between the directional CMTRG \cite{Corboz2014} and the symmetric CTMRG \cite{Baxter1968}. Here we choose the former, because it can be applied more generally. Both schemes are slightly different, also for symmetric tensors, because in the directional scheme the growing of the environment is done in an asymmetric way and the projectors are constructed from a product of $C$ tensors. Nonetheless, we find that these differences are very small, and that both give rise to the same expectation values.
\par In Fig.~\ref{fig:comp_energy} we have plotted the energy expectation value for the two optimized PEPSs as a function of the environment bond dimension $\chi$, evaluated with a boundary-MPS and a CTMRG environment. For a nearest-neighbour Hamiltonian ($J_2=0$) the energy can be evaluated by a single-row boundary-MPS or a one-by-two CTMRG environment; in the top panel of Fig.~\ref{fig:comp_energy} we observe that both approaches yield the same values. For a next-nearest-neighbour Hamiltonian ($J_2=1/2$) evaluating the energy from the boundary MPS requires a two-step process [Eq.~\eqref{eq:plaq}] or a two-by-two CTMRG environment [Eq.~\eqref{eq:ctmrg2}]. For small values of $\chi$ the boundary-MPS contraction is slightly more precise, but both methods clearly converge to the same value for larger $\chi$.
\par Note that the energy converges a lot slower for the frustrated PEPS, and therefore requires a larger value of $\chi$ for the same accuracy. This is in agreement with the result in Fig.~\ref{fig:lambda}.

\subsection{Structure factor}

\begin{figure}
\includegraphics[width=\columnwidth]{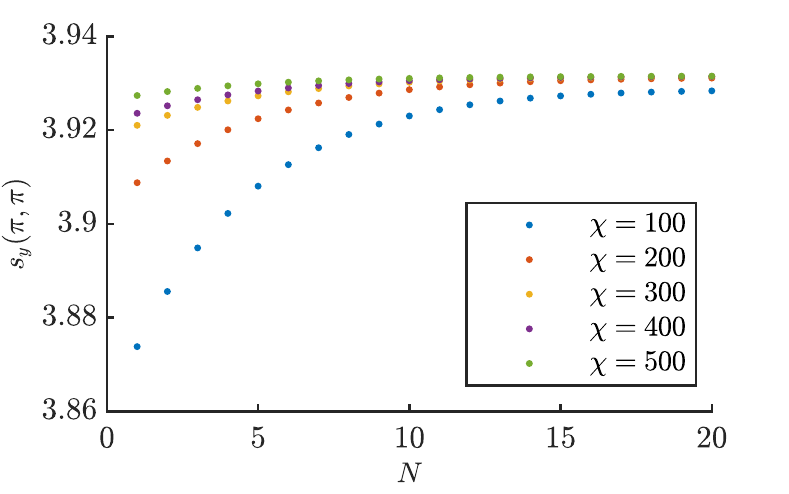}
\caption{The value of the structure factor as a function of window size, for different values of $\chi$.}
\label{fig:sf_windowsize}
\end{figure}

\begin{figure}
\subfigure{\includegraphics[width=\columnwidth]{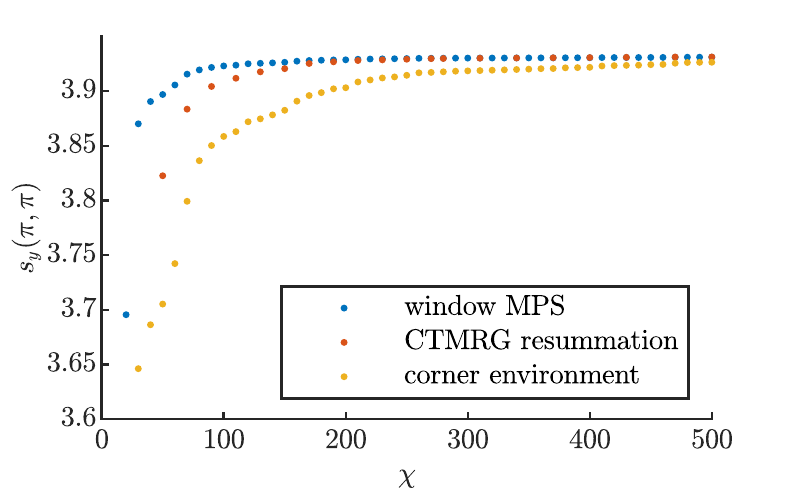}}
\subfigure{\includegraphics[width=\columnwidth]{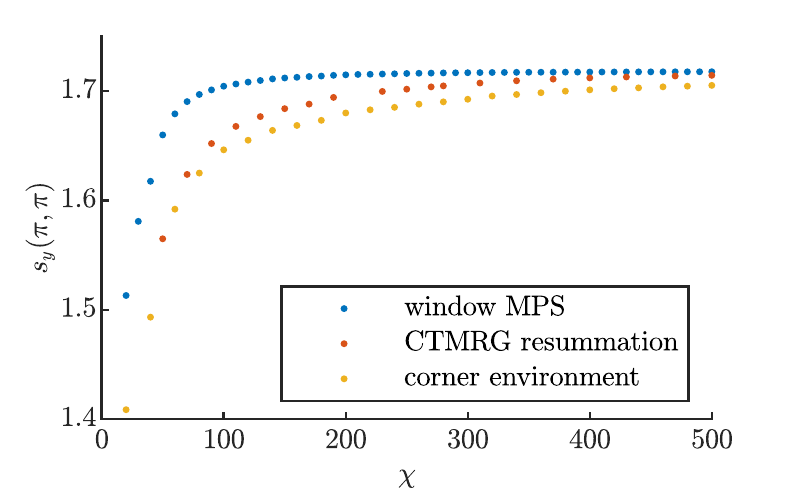}}
\caption{The value of the structure factor as a function of $\chi$, computed with the window-MPS approach with window size $N=10$ (blue), the CTMRG resummation scheme (red) and the channel environments (yellow). $J_2=0$ (top) and $J_2=1/2$ (bottom).}
\label{fig:sf_comp}
\end{figure}

\par Finally, we benchmark our new summation scheme based on the window-MPS approach to compute the static spin structure factor of the PEPS wavefunction
\begin{equation}
s_\alpha(\vec{k}) = \sum_{x,y} \e^{i(k_xx+k_yy)} \bra{\Psi(A)} S^\alpha_{(x,y)} S^\alpha_{(0,0)} \ket{\Psi(A)}.
\end{equation}
The window-MPS approach has two control parameters, the MPS bond dimension and the window size. In Fig.~\ref{fig:sf_windowsize}, we plot the value of the structure factor as a function of the window size, for five different values of $\chi$. We observe that a larger value of $N$ is needed to converge the result as $\chi$ is lower. This can be understood from the fact that a larger-$\chi$ MPS contains longer-range correlations, and that a local operation can influence the state over a larger region -- therefore, a small window can represent a larger perturbation when its bond dimension is higher. We also observe that the result for small $\chi$ does not converge to the correct result, as the initial boundary MPS does not carry the correct correlations -- this is analogous to simulating the time evolution of a local operator on a ground-state MPS with a too small bond dimension.
\par We can now compare our new summation scheme with the corner-environment approach \cite{Vanderstraeten2015} and CTMRG summation \cite{Corboz2016, Ponsioen2020}. Fig.~\ref{fig:sf_comp} shows that all three approaches converge to the same result, but with a clear difference in accuracy for smaller values of $\chi$. We observe that all schemes approach the converged value from below, but we have no good a priori reason to expect that this is always the case -- the window-MPS scheme takes variational steps, but the summed-up structure factor is not a variational quantity itself, in contrast to the norm of the PEPS wavefunction in Fig.\~ref{fig:lambda}. The window-MPS approach, which exploits variational principles through its evaluation of all contributions to the structure factor, is clearly the most accurate; for summing the contributions of each row, however, it requires us to perform a number of optimization sweeps. The CTMRG summation is less accurate, but only comes with little extra computational cost compared to the standard CTMRG scheme (one needs to keep track of separate environment tensors for the norm and the summed operator terms). The corner environment, finally, is the least accurate, but evaluating the structure factor just requires contracting a single one-dimensional tensor network and, therefore, comes at almost no cost on top of a boundary-MPS contraction. Choosing the summation scheme, therefore, depends a lot on the application, the required accuracy and the available computational resources.

\section{Discussion and outlook}

In this paper, we have identified a subclass of PEPS wavefunctions for which the transfer matrix is a Hermitian operator. This has allowed us to formulate the contraction as a variational problem, yielding an optimality condition that is independent of the algorithm that is actually used in practice. We have used this feature to compare state-of-the-art contraction algorithms such as the CTMRG and VUMPS algorithms, showing that they give comparable results for e.g. evaluating the energy of a PEPS. In addition, we have formulated a new scheme for computing general $N$-point functions, which is significantly more accurate as compared to existing alternative schemes. It would be instructive to compare our scheme to the generating-function approach of Ref.~\onlinecite{WeiLin2021}.
\par We believe that our new scheme will prove very useful in future PEPS methods and applications that go beyond ground-state properties. Indeed, a recent method for computing excitation spectra on top of a PEPS \cite{Vanderstraeten2015, Vanderstraeten2019b, Ponsioen2020, Ponsioen2021} relies heavily on summing two- and three-point functions. In addition, the time-dependent variational principle for uniform time evolution will require summing up two-point functions: the tangent vector that dictates the TDVP flow equations \cite{Vanderstraeten2019a} takes on a similar form as the energy gradient [see App.~\ref{sec:gradient}]. Finally, the idea of using window MPS in PEPS calculations will prove useful to simulate local dynamics against a uniform background -- simulating the response of the action of a local operator on the ground state for obtaining spectral functions.
\par For the Hermitian subclass of PEPS, we can unambiguously define the variationally optimal PEPS at a certain bond dimension $D$ and contracted with an environment bond dimension $\chi$. In Ref.~\onlinecite{Vanhecke2021}, this feature was exploited to formulate a scaling hypothesis in terms of a single effective length scale that results from both finite-$D$ and finite-$\chi$ effects.
\par Given the power of this subclass of PEPS, the question naturally arises under what conditions such a restriction can safely be imposed on the PEPS ansatz. One important restriction concerns the fact that the transfer matrix only has real eigenvalues. In the case of MPS, the eigenvalues of the transfer matrix are directly related to the wavevector of the dominant correlations in the state and, indirectly, to the minima of the quasiparticle dispersion relation on top of this state \cite{Zauner2015b}. This implies that a ground state with dominant incommensurate correlations is expected to be poorly described by our subclass of PEPS -- critical states with an incommensurate filling would satisfy this condition. It would be very interesting to track the limitations of the Hermitian subclass of PEPS for these systems, and what this implies for the contraction of their PEPS ground-state approximations.
\par In this paper we have only looked at PEPS on a square lattice, and it would be interesting to generalize the construction to other lattices such as the triangular or the kagome lattice. Also, we would like to investigate the case of highly non-trivial PEPS unit cells, which represent, e.g., stripe phases in the Hubbard model with a commensurate filling \cite{Zheng2017, Ponsioen2019} or complicated magnetization patterns \cite{Corboz2014b} in spin systems in an external magnetic field.

\begin{acknowledgments}
We would like to thank Juraj Hasik for inspiring discussions. This work was supported by  the Research Foundation Flanders, ERC grants QUTE (647905), ERQUAF (715861) and TENSORNETSIM (677061).
\end{acknowledgments}

\bibliography{./bibliography}

\appendix

\section{The energy gradient as a two-point function}
\label{sec:gradient}

In this Appendix we write out the expressions for evaluating the energy and the gradient for a given PEPS tensor, based on the summation scheme from Sec.~\ref{sec:summation}. We consider plaquette interactions on the square lattice. The energy expectation value is then given by
\begin{equation}
e = \diagram{a1_energy}{1},
\end{equation}
but evaluating this expression requires computing the fixed points of the two-row channel. We propose a different contraction technique, based on the window-MPS approach. We split up the Hamiltonian into two parts, and apply a first layer of the above diagram and approximate it by introducing a two-site tensor $P$ as
\begin{multline}
\diagram{a1_energy}{2} \\ \approx \diagram{a1_energy}{3}.
\end{multline}
Finding the optimal tensor $P$ is straightforward since the cost function is linear in $P$. We find
\begin{equation}
\diagram{a1_energy}{4} = \diagram{a1_energy}{5}.
\end{equation}
The energy expectation value is then obtained as
\begin{equation} \label{eq:plaq}
e = \diagram{a1_energy}{6}.
\end{equation}

The gradient is then computed in a similar fashion as the structure factor. Indeed, the gradient for the energy expectation value
\begin{align}
g &= 2 \partial_{\bar{A}} \frac{\bra{\Psi(\bar{A})} H \ket{\Psi(A)}}{\braket{\Psi(\bar{A}) | \Psi(A)}} \\
&= 2 \frac{ \bra{\partial_{\bar{A}} \Psi(\bar{A})} H - e(A,\bar{A}) \ket{\Psi(A)}} {\braket{\Psi(\bar{A}) |\Psi(A)}}
\end{align}
is an infinite sum where in each term one differentiates with respect to one tensor in the bra layer. The first terms are the ones where the differential is on the Hamiltonian
\begin{equation}
g_0 = \diagram{a1_gradient}{1}.
\end{equation}
Then we continue in the same row, with the term
\begin{equation}
g_1 = \diagram{a1_gradient}{2} \left ( 1 -\diagram{a1_gradient}{3} \right)^{-1} \diagram{a1_gradient}{4}.
\end{equation}
Finally we have all the terms of the form
\begin{equation}
g_2 = \diagram{a1_gradient}{5},
\end{equation}
which can, again, be computed similarly to an out-of-time correlator using a finite-window MPS. Indeed, denoting the Hamiltonian MPS as
\begin{equation}
\ket{\Psi_H(M)} = \diagram{a1_gradient}{6}
\end{equation}
and the MPS with the differential
\begin{multline}
\ket{\Psi_\partial(M)} = \diagram{a1_gradient}{7} \\ + \diagram{a1_gradient}{8} + \dots,
\end{multline}
we find
\begin{align}
g_2 &= \sum_{n=1}^\infty \bra{\Psi_\partial(M)} T^{n-1} \ket{\Psi_H(M)} \\
& = \bra{\Psi_\partial(M)} (1-T)^{-1} \ket{\Psi_H(M)}.
\end{align}
This correlator is most easily computed by a finite-window MPS, i.e., we sequentially apply the transfer matrix and approximate the state as
\begin{equation}
T^n \ket{\Psi_H(M)} \approx \ket{\Psi(M;\{N_i\})},
\end{equation}
and compute the overlaps
\begin{equation}
\braket{\Psi_\partial(M) | \Psi(M;\{N_i\})} .
\end{equation}

\section{Variational characterization of boundary MPS methods for multi-row transfer matrices}
\label{sec:multivumps}

In this Appendix we give a detailed explanation of variational principles and contraction routines for larger unit cells. Such an explanation is intended to complement the discussion in Ref.~\onlinecite{Nietner2020}, where it was missed that the multi-site VUMPS algorithm might not converge for unit cells with more than two rows.

\subsection{Discussion}

\par The VUMPS fixed point equations [Eqs.~\eqref{eq:vumps1}-\eqref{eq:vumps4}] are derived through the projection of the eigenvalue equation Eq.~\eqref{eq:eigEq} onto the tangent space to the manifold of MPS,
\begin{equation} \label{eq:eigEqTangent}
	\cP_M \left (T \ket{\Psi_M} - \Lambda \ket{\Psi_M} \right ) = 0 \,,
\end{equation}
where $ \cP_M $ is the projector onto the tangent space of the MPS manifold at the point parameterized by $ M $. For the case of a transfer matrix with a one-by-one unit cell where the tensor parameterizing $ T $ is Hermitian, the solution of Eq.~\eqref{eq:eigEqTangent} directly corresponds to the variational optimum of a real-valued objective function representing the free-energy density Eq.~\eqref{eq:costFun}. Thus, the VUMPS algorithm for a transfer matrix  with a one-by-one Hermitian unit cell gives an efficient method for obtaining the approximate leading eigenvector in a variationally optimal way.
\par For transfer matrices with a non-trivial unit cell, the cost of a naive application the VUMPS algorithm would  scale exponentially with the size of the unit cell. It is therefore desirable to work with a variational algorithm that directly takes into account the non-trivial structure of the unit cell. One such algorithm is the multi-site generalization of the VUMPS algorithm put forward in Ref.~\onlinecite{Nietner2020}. In this approach, one finds the approximate boundary MPS for an $ n $-row Hermitian transfer matrix $ T = T_n \dots T_2 T_1 $, where $ T_i^\dagger =  T_{n-i+1} $, in the following way. Instead of directly finding a single MPS $ \ket{\Psi_{\{M\}}} $ which satisfies $ T \ket{\Psi_{\{M\}}} = \Lambda \ket{\Psi_{\{M\}}} $, one formulates a recursive eigenvalue problem in terms of a set of MPSs $ \ket{\Psi_{\{M_i\}}} $ that satisfy
\begin{align}
	T_1 \ket{\Psi_{\{M_1\}}} &= \Lambda_1 \ket{\Psi_{\{M_2\}}} \,, \nonumber \\
	T_2 \ket{\Psi_{\{M_2\}}} &= \Lambda_2 \ket{\Psi_{\{M_3\}}} \,, \nonumber \\
	&\cdots \nonumber \\
	T_n \ket{\Psi_{\{M_n\}}} &= \Lambda_n \ket{\Psi_{\{M_1\}}} \,, \label{eq:multiEigEq}
\end{align}
where $ \Lambda = \prod_{i = 1}^{n} \Lambda_i $. The appropriate fixed point equations for a multi-row boundary MPS algorithm are obtained by projecting each of these equations onto the tangent space at the appropriate point on the MPS manifold,
\begin{equation} \label{eq:multiEigEqTangent}
	\cP_{\{M_{i+1}\}} \left (T_i \ket{\Psi_{\{M_i\}}} - \Lambda_i \ket{\Psi_{\{M_{i+1}\}})}\right ) = 0 \,.
\end{equation}
Whereas in the case of a single-row transfer matrix this tangent-space criterion is in direct correspondence with the variational optimum of a real-valued cost function, this is no longer necessarily the case when dealing with larger unit cells. For a two-row transfer matrix there is no issue, as the conditions [Eqs.~\eqref{eq:multiEigEqTangent}] still correspond to the variational optimum of Eq.~\eqref{eq:var_mps2}. However, for transfer matrices with three or more rows this characterization in terms of optimizing the free-energy density breaks down, as mentioned in Sec.~\ref{sec:vumps_mps3}.
\par The conditions [Eqs.~\eqref{eq:multiEigEqTangent}] can however still be characterized in terms of an optimality criterion: they are in direct correspondence with the variational optimum of the normalized fidelity between consecutive MPS. Specifically, if
\begin{equation} \label{eq:multiSiteFidelity}
    \{M_{i+1}\} = \argmax_{\{M_{i+1}\}} \frac{ \left|\bra{\Psi_{\{M_{i+1}\}}} T_i \ket{ \Psi_{\{M_i\}}} \right|^2} { \braket{\Psi_{\{M_{i+1}\}}|\Psi_{\{M_{i+1}\}}} },
\end{equation}
then Eq.~\eqref{eq:multiEigEqTangent} must hold. This variational characterization in terms of the normalized fidelity naturally leads to a multi-row boundary MPS algorithm in the form of a power method, as originally described in section A.2 of Ref.~\onlinecite{Nietner2020}. Start with some initial MPS $ \ket{\Psi_{\{M_1^{(0)}\}}} $ and consecutively apply the individual transfer matrix layers $ T_i $, each time approximating the resulting MPS by some other MPS of a given bond dimension by variationally optimizing the normalized fidelity Eq.~\eqref{eq:multiSiteFidelity} \cite{Vanhecke2021b}. This procedure is repeated until convergence is reached, where the result will then correspond to the fixed point characterized by [Eqs.~\eqref{eq:multiEigEq}]. In Sec.~\ref{sec:2dIsingNumerics} we argue that this different variational characterization of the multi-row algorithm in terms of an optimal normalized fidelity is compatible with the characterization in terms of optimizing the free-energy density, for systems that are sufficiently gapped, by considering the example of the two-dimensional classical Ising model. Note that the computational complexity of this power method is again linear in both linear dimensions of the unit cell $L_x$ and $L_y$.
\par The algorithm proposed in Ref.~\onlinecite{Nietner2020}, however, goes beyond this power method. The multi-site VUMPS algorithm does not optimize [Eqs.~\eqref{eq:multiSiteFidelity}] sequentially for one MPS at a time, but in fact iteratively solves a global fixed-point equation which is obtained by writing down all eigenvalue equations [Eqs.~\eqref{eq:multiEigEq}] in their tangent space projected form simultaneously, thereby optimizing over all MPS at once. This gives rise to a more efficient algorithm compared to the simple power method. As the conditions [Eqs.~\eqref{eq:multiEigEqTangent}] are contained within the solution of this global fixed point equation, if the multi-site VUMPS algorithm converges then the result is guaranteed to be optimal in the sense of the normalized fidelity.
\par An important point, however, is that the multi-site algorithm may not converge for certain transfer matrices with three or more rows. This is caused by the fact that the algorithm optimizes over all MPS at the same time in order to solve a single global fixed point equation, instead of optimizing each fidelity individually as is actually implied by the tangent space conditions [Eqs.~\eqref{eq:multiEigEqTangent}]. As stated in Sec.~\ref{sec:vumps_mps3}, this phenomenon is related to the fact that the multi-site VUMPS fixed point equations can actually be related to an invalid variational principle of the form Eq.~\eqref{eq:var_mps3} in this case. So while the more efficient multi-site VUMPS algorithm usually converges in which case the obtained solution automatically satisfies [Eqs.~\eqref{eq:multiEigEqTangent}] in an optimal way, the algorithm can fail in which case the more robust power method should be used to obtain the desired fixed point.

\subsection{Numerical results for the 2d Ising model}
\label{sec:2dIsingNumerics}

\par We now provide some numerical results in support of this discussion. The model we study is the classical ferromagnetic Ising model on the infinite two-dimensional square lattice with Hamiltonian
\begin{equation} \label{eq:isingHam}
	H = -J \sum_{\braket{i,j}} s_i s_j \,.
\end{equation}
The corresponding partition function
\begin{equation} \label{eq:isingPartFun}
	\cZ = \sum_{\{s_i\}} \txe^{-\beta H}
\end{equation}
can be written as the contraction of a 2-D tensor network,
\begin{equation} \label{eq:isingPartFunTn}
	\cZ = \diagramScale{a2_ising}{1}{.45}\;,
\end{equation}
where the black dots represent $ \delta $-tensors
\begin{equation} \label{eq:deltaTensor}
	\diagramScale{a2_ising}{2}{.45} = \begin{cases}
	1 \quad \text{if } i = j = k = l \,,\\
	0 \quad \text{otherwise,}
	\end{cases}
\end{equation}
and Boltzmann weights
\begin{equation} \label{eq:boltzmannWeights}
    \diagramScale{a2_ising}{3}{.45} = \begin{pmatrix}
	\txe^{\beta J} && \txe^{-\beta J} \\
	\txe^{-\beta J} && \txe^{\beta J}
	\end{pmatrix}
\end{equation}
are place on each bond. Instead of the conventional way of simulating this model using a single-row transfer matrix, we will resort to a three-row transfer matrix here. Our approach is inspired by problems for which the use of a two-by-two unit cell transfer matrix is required, either explicitly or as a result of breaking translational invariance. For these models it often occurs that, while the total partition function has reflection and time-reversal symmetry, the corresponding transfer matrix unit cell is non-Hermitian and the variational characterization of boundary MPS contraction is therefore not valid. This occurs for example for certain classical lattice gauge theories in two spatial dimensions. In such cases, the partition function can often be transformed into an equivalent partition function with a larger Hermitian three-by-three unit cell by means of an identity for $ \delta $-tensors, which in the current case reads
\begin{equation} \label{eq:sqrtTrick}
    \diagramScale{a2_ising}{7}{.45} = \diagramScale{a2_ising}{8}{.45}.
\end{equation}
Here the black dots on the far left and right of the right hand side represent three-leg $ \delta $-tensors just as in Eq.~\eqref{eq:deltaTensor} and
\begin{equation} \label{eq:boltzmannWeightsSqrt}
    \diagramScale{a2_ising}{15}{.45} = \begin{pmatrix}
	\txe^{\frac{\beta J}{2}} && \txe^{-\frac{\beta J}{2}} \\
	\txe^{-\frac{\beta J}{2}} && \txe^{\frac{\beta J}{2}} 
	\end{pmatrix} \;.
\end{equation}
Applied to Eq.~\eqref{eq:isingPartFunTn}, this gives rise to a partition function with a Hermitian three-by-three unit cell, indicated by the dashed square,
\begin{equation} \label{eq:isingPartFunTnSymmetrized}
	\cZ = \diagramScale{a2_ising}{9}{.45} .
\end{equation}
If we absorb the bond matrices into the sites symmetrically this gives a Hermitian three-row transfer matrix,
\begin{equation} \label{eq:isingTransferSymmetrized}
	T = \diagramScale{a2_ising}{16}{.35} ,
\end{equation}
where the bond and plaquette tensors of the unit cell are defined as
\begin{equation} \label{eq:symmetrizedTensorsDef}
    \diagramScale{a2_ising}{13}{.35} = \diagramScale{a2_ising}{14}{.45} , \quad
    \diagramScale{a2_ising}{11}{.35} = \diagramScale{a2_ising}{12}{.45} .
\end{equation}
Given this three-row transfer matrix, we now illustrate the points made in the above discussion.
\par We approximate the leading eigenvector of Eq.~\eqref{eq:isingTransferSymmetrized} as a boundary MPS by directly using multi-site algorithms, as well as by blocking the unit cell and employing the single-site VUMPS algorithm. In the following we will use the term \enquote{single-row error} to denote the deviation of the boundary MPS with respect to the constraint Eq.~\eqref{eq:eigEqTangent} where all rows of the transfer matrix are blocked, indicating its distance from the variational optimum of the free-energy density Eq.~\eqref{eq:var_mps1}. The term \enquote{multi-row error} will be used to denote the deviation with respect to the set of constraints [Eqs.~\eqref{eq:multiEigEqTangent}], indicating its distance from the variational optimum of the normalized fidelity [Eqs.~\eqref{eq:multiSiteFidelity}]. We set $ J = 1 $ in Eq.~\eqref{eq:isingHam} and consider a temperature range around the critical temperature $ T_c = 2 / \log(1 + \sqrt{2})) $. Using a bond dimension of $ \chi = 50 $ for the boundary MPS, we apply the power method described above, the multi-site VUMPS algorithm \cite{Nietner2020} as well as the single-site VUMPS algorithm for the blocked unit cell \cite{Fishman2018} and analyze the results.
\par
\begin{figure}
\subfigure{\includegraphics[width=\columnwidth]{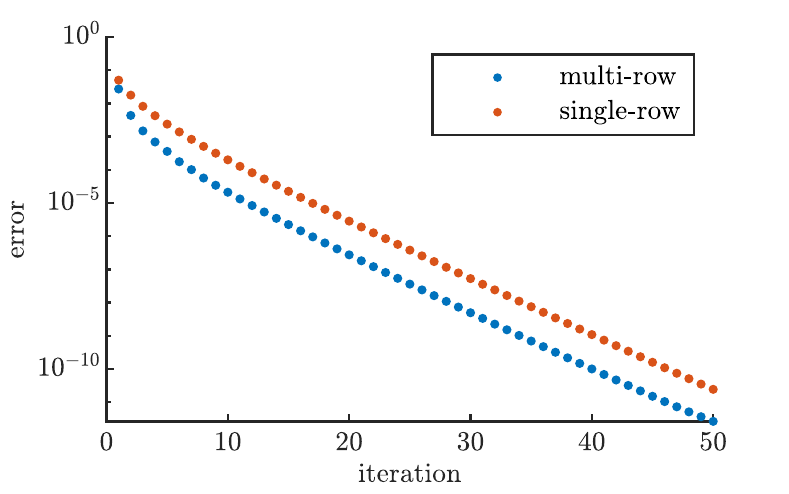}}
\subfigure{\includegraphics[width=\columnwidth]{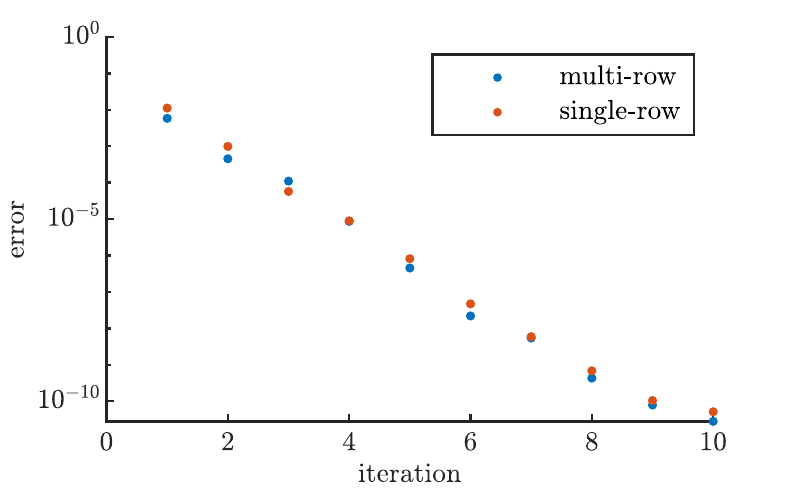}}
\caption{Multi- and single-row error at each iteration using the power method (top) and the multi-site VUMPS algorithm (bottom) for determining the boundary MPS of Eq.~\eqref{eq:isingTransferSymmetrized}, at $ T = 0.95 \, T_c $.}
\label{fig:multiIsingSymm_multi_single_comparison}
\end{figure}

\begin{figure}
\includegraphics[width=\columnwidth]{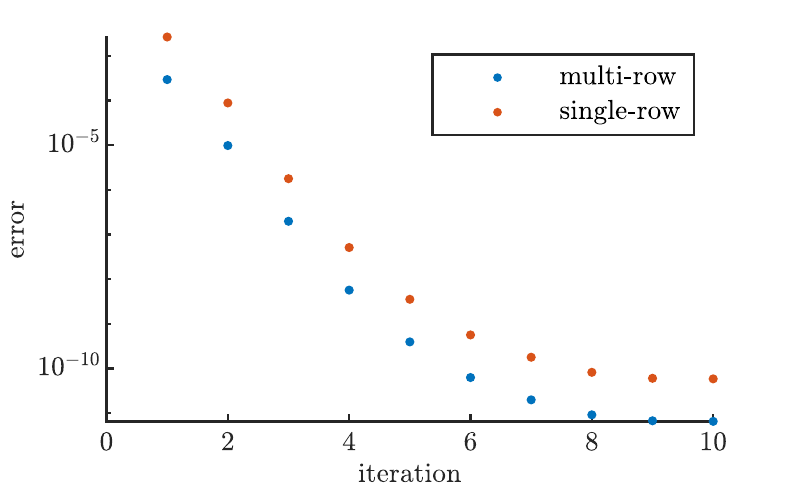}
\caption{Multi- and single-row error at each iteration using the single-site VUMPS algorithm for determining the boundary MPS of \eqref{eq:isingTransferSymmetrized} when blocking all layers, at $ T = 0.95 \, T_c $.}
\label{fig:multiIsingSymm_single_multi_comparison}
\end{figure}

 In order to motivate the validity of the variational characterization in terms of an optimal fidelity per site, we take the MPS obtained with the multi-row power method and the multi-site VUMPS algorithm at each iteration and use it to compute the single-row error with respect to the blocked transfer matrix. The results are shown in Fig.~\ref{fig:multiIsingSymm_multi_single_comparison}, for $ T = 0.95 \, T_c $. The superior efficiency of the multi-site VUMPS algorithm over the power method is immediately apparent here. 
Conversely, we take the MPS obtained with the single-site VUMPS algorithm for the blocked transfer matrix and use it to compute the multi-row error with respect to the multi-row transfer matrix by splitting the state into a three-site unit cell MPS by means of SVD. The result is shown in Fig.~\ref{fig:multiIsingSymm_single_multi_comparison} for $ T = 0.95 \, T_c $.
We see that both variational characterizations are compatible at a sufficient distance from criticality, i.e., if one characterization is used to optimize the boundary MPS then the other characterization is automatically optimized as well.

\par
\begin{figure}
\subfigure{\includegraphics[width=\columnwidth]{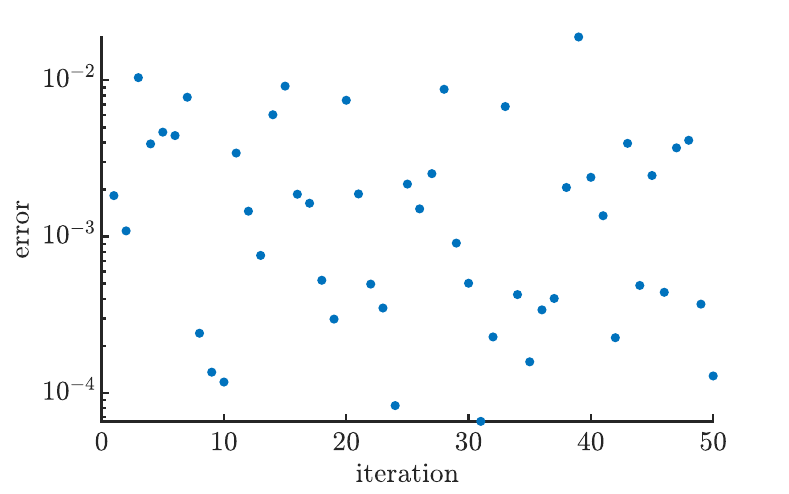}}
\subfigure{\includegraphics[width=\columnwidth]{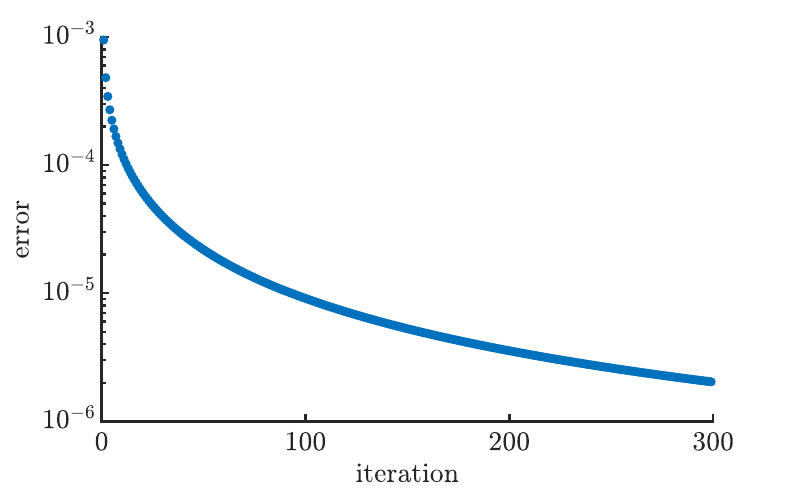}}
\caption{Multi-row error at each iteration using the multi-site VUMPS algorithm (top) and the power method (bottom) for determining the boundary MPS of \eqref{eq:isingTransferSymmetrized} at $ T = T_c $.}
\label{fig:multiIsingSymm_critical}
\end{figure}

 As stated above, the multi-site VUMPS algorithm \cite{Nietner2020} can suffer from stability issues for certain transfer matrices with three or more rows. In order to illustrate this, we monitor the multi-row error for the first 50 iterations when determining the boundary MPS of Eq.~\eqref{eq:isingTransferSymmetrized} at $ T = T_c $ using both the multi-site VUMPS algorithm and the multi-row power method. The result is shown in Fig.~\ref{fig:multiIsingSymm_critical}. It can clearly be seen from the top panel that the behavior of the deviation with respect to Eq.~\eqref{eq:multiEigEqTangent} is entirely erratic after a few iterations of the multi-site VUMPS algorithm: the algorithm becomes unstable and will not converge. In contrast, this behavior does not occur when using the power method: the error measure keeps decreasing with the number of iterations and the algorithm will converge.

\par
\begin{figure}
\includegraphics[width=\columnwidth]{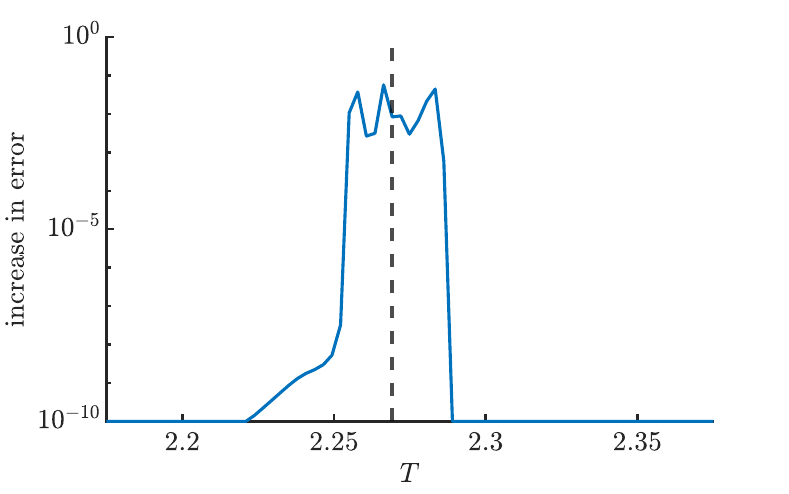}
\label{fig:multiIsingSymm_vumpsStability_chi_50}
\caption{Increase of the multi-site error after 40 additional multi-site VUMPS iterations starting from the power method result for a temperature range $ (.95:.00125:1.05) \cdot T_c $. The dashed vertical line indicates $ T = T_c $. Differences smaller than $ 10^{-10} $ are ignored since these are tainted by numerical noise.}
\label{fig:vumpsStability}
\end{figure}

 To further illustrate this instability, we take the fixed point obtained after 200 iterations of the power method and use this as the input for the multi-site VUMPS algorithm. The increase of the deviation with respect to [Eqs.~\eqref{eq:multiEigEqTangent}] for the multi-site VUMPS result after 40 additional iterations compared to the power method result is shown in Fig.~\ref{fig:vumpsStability}. It can clearly be seen that the multi-site VUMPS algorithm becomes unstable in a finite temperature region around the critical temperature. Our simulations suggest that the size of this window is largely independent of the MPS bond dimension, while its position shifts very slightly towards lower temperatures when the bond dimension is increased.

\end{document}